\documentclass[preprint,12pt]{elsarticle}

\usepackage{amssymb}
\usepackage{amsmath}
\usepackage{booktabs}
\usepackage{longtable}
\usepackage{pdflscape}
\usepackage{tabularx}
\usepackage{graphicx}
\usepackage{subfig}
\usepackage{caption}
\usepackage{hyperref}
\hypersetup{colorlinks,allcolors=black}
\usepackage{cleveref}
\usepackage{siunitx}
\usepackage{acronym}
\usepackage{xurl}

\usepackage{lineno}

\journal{}

\begin{document}

% defining new column which can take width as an argument
\newcolumntype{L}[1]{>{\raggedright\arraybackslash}p{#1}}

\begin{frontmatter}



\title{Modelling Regional Solar Photovoltaic Capacity in Great Britain}

\author[label1,label2]{Alghanem, H.\corref{cor1}} \ead{halghanem1@sheffield.ac.uk}

\affiliation[label1]{organization={The University of Sheffield},
            addressline={Department of Physics and Astronomy},
            city={Sheffield},
            postcode={S3 7RH},
            country={United Kingdom}}

\affiliation[label2]{organization={Imam Abdulrahman Bin Faisal University},
            addressline={Department of Physics, College of Science and Humanities-Jubail},
            city={Dammam},
            postcode={34212},
            country={Saudi Arabia}}

\author[label1]{Buckley, A.} \ead{Alastair.Buckley@sheffield.ac.uk} %% Author name

\cortext[cor1]{Corresponding author}

% \affiliation{organization={University of Sheffield},%Department and Organization
%             addressline={Hicks Building}, 
%             city={Sheffield},
%             country={UK}}

%% Abstract
\begin{abstract}

Great Britain aims to meet growing electricity demand and achieve a fully decarbonised grid by 2035, targeting 70 GW of solar photovoltaic (PV) capacity. However, grid constraints and connection delays hinder solar integration. To address these integration challenges, various connection reform processes and policies are being developed \cite{national_energy_system_operator_neso_connections_nodate}. This study supports the connection reforms with a model that estimates regional PV capacity at the NUTS 3 level, explaining 89\% of the variation in capacity, with a mean absolute error of 20 MW and a national mean absolute percentage error of 5.4\%. Artificial surfaces and agricultural areas are identified as key factors in deployment. The model has three primary applications: disaggregating national PV capacity into regional capacity, benchmarking regional PV deployment between different regions, and forecasting future PV capacity distribution. These applications support grid operators in generation monitoring and strategic grid planning by identifying regions where capacity is likely to be concentrated. This can address grid connection delays, plan network expansions, and resolve land-use conflicts.

\end{abstract}


% \begin{highlights}
% \item Identified key geographical factors influencing solar PV deployment in European regions. 
% \item Developed models for estimating solar PV capacity at the NUTS 2 level across 36 European countries.
% \item Provided a dataset of regional PV capacity estimates for 333 European regions, covering the period from 2010 to 2023.
% \item Developed a regional Solar PV Deployment Index to assess and compare PV capacity across regions in Europe.
% \end{highlights}

%% Keywords
% \begin{keyword}
% % keywords here, in the form: keyword \sep keyword

% solar PV Capacity \sep Regional PV \sep PV Capacity Forecasting \sep Benchmarking PV Deployment \sep Capacity Disaggregation

% %% PACS codes here, in the form: \PACS code \sep code

% %% MSC codes here, in the form: \MSC code \sep code
% %% or \MSC[2008] code \sep code (2000 is the default)

% \end{keyword}

\end{frontmatter}

% \section*{Highlights}

% \begin{itemize}
%     \item Identified key geographical factors influencing solar photovoltaic deployment in the NUTS 3 regions of Great Britain.
%     \item Identified artificial surfaces and agricultural areas as key factors in solar photovoltaic deployment.
%     \item Developed models for estimating solar PV capacity at the NUTS 3 level across 168 regions in Great Britain.
%     \item Provided a dataset of regional photovoltaic capacity estimates for 168 Great Britain regions, covering the period from 2010 to 2023.
%     \item Developed a regional Solar Photovoltaic Deployment Index to assess and compare PV capacity across regions in Great Britain.
% \end{itemize}

% \begin{itemize}
%     % \item Identified key geographical factors influencing solar PV deployment in NUTS 3 regions.
%     \item Identified key geographical factors influencing regional solar PV deployment in GB.
%     \item Identified artificial surfaces and agricultural areas as key factors in PV deployment.
%     \item Developed models for estimating solar PV capacity at the NUTS 3 level across 168 regions.
%     \item Provided a dataset of regional PV capacity estimates covering the years 2010 to 2023.
%     \item Developed a regional Solar Photovoltaic Deployment Index.
% \end{itemize}

% \linenumbers


\section*{Keywords}
Great Britain Solar PV, Solar PV Capacity, Regional PV, PV Capacity Forecasting, Benchmarking PV Deployment, Capacity Disaggregation.

% \section*{Word count: }

% \section*{List of Abbreviations}
% \begin{acronym}
%     \acro{PV}{Photovoltaic}
%     \acro{MW}{Megawatt}
%     \acro{SHAP}{SHapley Additive exPlanations}
%     \acro{SPVDI}{Solar PV Deployment Index}
%     \acro{MSE}{Mean Squared Error}
%     \acro{RMSE}{Root Mean Squared Error}
%     \acro{MAE}{Mean Absolute Error}
%     \acro{MAPE}{Mean Absolute Percentage Error}
%     \acro{EU}{European Union}
%     \acro{NUTS}{Nomenclature of Territorial Units for Statistics}
%     \acro{GHI}{Global Horizontal Irradiance}
%     \acro{GDP}{Gross Domestic Product}
%     \acro{GERD}{Gross Expenditure on Research and Development}
%     \acro{$R^2$}{Coefficient of Determination}
%     \acro{$GHI$}{Global Horizontal Irradiance}
%     \acro{$IRENA$}{International Renewable Energy Agency}
%     \acro{GLMs}{General linear models}  
% \end{acronym}

\section*{List of Abbreviations}
\begin{acronym}
    % General terms
    % \acro{EU}{European Union}
    % \acro{GDP}{Gross Domestic Product}
    % \acro{GERD}{Gross Expenditure on Research and Development}
    
    % Solar PV-specific terms
    \acro{PV}{Photovoltaic}
    % \acro{SPVDI}{Solar PV Deployment Index}
    % \acro{IRENA}{International Renewable Energy Agency}

    % Statistical and modeling terms
    \acro{SHAP}{Shapley Additive Explanations}
    \acro{XGBoost}{Extreme Gradient Boosting}
    \acro{GLMs}{General Linear Models}
    \acro{PCA}{Principal Component Analysis}
    \acro{MSE}{Mean Squared Error}
    \acro{RMSE}{Root Mean Squared Error}
    \acro{MAE}{Mean Absolute Error}
    \acro{MAPE}{Mean Absolute Percentage Error}
    \acro{$R^2$}{Coefficient of Determination}

    % Regional and classification terms
    \acro{NUTS}{Nomenclature of Territorial Units for Statistics}
    \acro{GDPR}{General Data Protection Regulation}
    \acro{DNO}{Distribution Network Operator}
    \acro{GVA}{Gross Value Added}
    \acro{ROCS}{Renewable Obligation Certificates}
    \acro{ofgem}{Office of Gas and Electricity Markets}
    \acro{ITL}{International Territorial Level}
    \acro{GDHI}{Gross Disposable Household Income}

    % Energy-related terms
    \acro{MW}{Megawatt}
    \acro{GHI}{Global Horizontal Irradiance}
\end{acronym}

\section{Introduction}
\label{introduction}

The demand for electricity in Great Britain is projected to grow significantly, potentially increasing by 65\% by 2035. To address this rising demand, the UK government has set ambitions goals including achieving a fully decarbonised electricity system \cite{national_energy_system_operator_neso_beyond_2024} and installing 70 GW of solar power by 2035 \cite{felicia_rankl_planning_2024}. 

Significant progress has already been made, with 2023 marking the first year that renewable energy generation surpassed fossil fuel generation in Britain. However, the electricity grid faces mounting challenges as it nears its capacity limits. Investment in renewable generation has outpaced spending on transmission infrastructure over the past decade, creating constraints that restrict the grid’s ability to transport electricity. Consequently, energy is sometimes wasted when renewable sources are curtailed to prevent grid overloading \cite{national_energy_system_operator_neso_beyond_2024}.

A major obstacle to scaling up renewable energy, including solar PV, is grid connection delays. As of October 2024, 732 GW of projects were queued to connect to Great Britain’s transmission network, with renewables accounting for approximately 363 GW \cite{energy_networks_association_ofgem_2024}. The queue grows as investors hedge their bets on where grid connection approvals might be granted, submitting multiple speculative applications for the same or similar projects in different locations. This strategy is often employed to improve the likelihood of securing a connection in a system where the timing and location of approvals are uncertain. However, this approach leads to an artificially inflated connection queue, as many of these projects are unlikely to materialize. It also creates inefficiencies for grid operators, who must process and manage a large volume of speculative applications, slowing down the approval process for viable projects \cite{uk_government_clean_2024}. These delays are further exacerbated by insufficient physical network infrastructure, such as cables, transformers, and substations, which are critical for accommodating new connections \cite{environmental_audit_committee_technological_2023,exawatt_written_nodate}. This reflects the growing strain on Great Britain’s grid in the context of a rapidly decarbonizing energy system, a challenge mirrored globally as countries transition from fossil fuels to renewable energy sources \cite{electricity_system_operator_eso_connections_2023}.

To tackle these challenges, the UK has introduced the Connections Reform initiative to improve the efficiency of the grid connection process \cite{electricity_system_operator_eso_connections_2023,uk_government_clean_2024}. Historically, this process has been reactive, addressing individual connection requests with little consideration of the broader network needs. The reform aims to streamline the process, prioritize projects nearing completion, and enhance transparency for developers. However, these measures alone cannot resolve the underlying need for a comprehensive, long-term strategy to align grid development with future energy demand and renewable deployment \cite{department_for_energy_security__net_zero_desnz_open_2024,environmental_audit_committee_enabling_2024}.

Strategic planning is especially important for solar PV deployment. Limited data on installations, due to General Data Protection Regulation (GDPR) and commercial data restrictions, complicate efforts to plan and invest in grid infrastructure effectively \cite{environmental_audit_committee_technological_2023}, and reduces the accuracy of PV generation estimates \cite{exawatt_written_nodate}. The lack of Distribution Network Operator (DNO) targets for new PV grid connections is also a barrier to reaching the UK's target of 70 GW by 2035 \cite{exawatt_written_nodate}. Moreover, accurate regional capacity modelling is critical for reducing errors in PV generation estimates. While the national capacity error is around 5\% \cite{huxley_uncertainties_2022}, regional errors are likely higher given that aggregation reduces the error.

Developing a model to estimate regional PV capacity based on geographical factors (social, economic, land use, and climatic) could address these challenges by proactively identifying regions where grid development might be needed. The model could help with setting realistic PV connection targets for DNOs and provide more accurate estimates of regional PV capacity, thereby improving generation monitoring. A regional PV capacity model could help policymakers identify areas where deployment lags behind expectations, enabling targeted interventions to support underserved regions. 

While we focus on the GB system in this paper, the issues are shared amongst countries. For example, grid connection constraints present significant challenges in various countries, including Austria \cite{gajdzik_barriers_2023}, Bulgaria \cite{gajdzik_barriers_2023}, Croatia \cite{christina_prifti_grid_2024}, Chile \cite{nasirov_investors_2015}, Finland \cite{christina_prifti_grid_2024}, France \cite{christina_prifti_grid_2024}, Germany \cite{christina_prifti_grid_2024}, Greece \cite{christina_prifti_grid_2024,gajdzik_barriers_2023}, Hungary \cite{gajdzik_barriers_2023}, Ireland \cite{christina_prifti_grid_2024}, Italy\cite{christina_prifti_grid_2024}, Netherlands \cite{zsuzsanna_pato_gridlock_2024,christina_prifti_grid_2024}, Poland \cite{christina_prifti_grid_2024}, Spain \cite{christina_prifti_grid_2024,ren21_renewables_2023}, Sweden \cite{christina_prifti_grid_2024}, Turkey \cite{christina_prifti_grid_2024}, and the United States \cite{gorman_grid_2024}.

% While we focus on the GB system in this paper, the issues are shared amongst countries. For example, grid connection constraints present significant challenges in various countries, including Austria \cite{gajdzik_barriers_2023}, Bulgaria \cite{gajdzik_barriers_2023}, Croatia \cite{christina_prifti_grid_2024}, Finland \cite{christina_prifti_grid_2024}, France \cite{christina_prifti_grid_2024}, Germany \cite{christina_prifti_grid_2024}, Greece \cite{christina_prifti_grid_2024,gajdzik_barriers_2023}, Hungary \cite{gajdzik_barriers_2023}, Ireland \cite{christina_prifti_grid_2024}, Italy\cite{christina_prifti_grid_2024}, Netherlands \cite{zsuzsanna_pato_gridlock_2024,christina_prifti_grid_2024}, Poland \cite{christina_prifti_grid_2024}, Spain \cite{christina_prifti_grid_2024,ren21_renewables_2023}, Sweden \cite{christina_prifti_grid_2024}, Turkey \cite{christina_prifti_grid_2024}, the United States \cite{gorman_grid_2024}, and Chile \cite{nasirov_investors_2015}. 

\subsection{Literature Review}
Addressing grid connection constraints requires a deeper understanding of the geographical factors that influence solar PV deployment. Previous studies have highlighted a range of key factors that impact deployment, including social, economic, climatic, and land use variables. Population is positively correlated with solar PV deployment at the country level globally \cite{alghanem_global_2024}. This is expected, as countries with larger populations generally require more electricity, which drives higher solar PV installation rates. However, the relationship between solar PV deployment and population becomes more complex when examined at regional and subregional scales. For example, studies have found a negative correlation between solar PV deployment and population at the regional level in both Germany \cite{mayer_deepsolar_2020} and the UK \cite{balta-ozkan_regional_2015} and at the subregional level in Australia \cite{fuentes_solar_2024}. In the United States, \citet{yu_deepsolar_2018} observed that residential solar PV deployment peaks at a population density of 1,000 people per square mile. Areas with very high population densities tend to have lower levels of small-scale residential solar PV deployment, as urban environments often lack suitable rooftops for installations. Conversely, regions with medium population densities are more likely to have higher solar PV capacities due to the prevalence of detached houses with rooftops that are ideal for solar PV systems. 

Education is positively correlated with solar PV deployment at the country level globally \cite{alghanem_global_2024}, at the regional level in the UK \cite{balta-ozkan_regional_2015}, and at the subregional level in Connecticut, USA \cite{yu_deepsolar_2018}, and England \cite{laura_williams_identifying_2012}. Countries with higher education levels tend to have stronger economies, which often translates into higher electricity consumption and greater investment in solar PV installations. Furthermore, higher levels of education are associated with increased environmental awareness and pro-environmental behaviour \cite{meyer_does_2015,wang_green_2022,ozbay_exploring_2022}, which can further drive solar PV adoption.

The influence of neighbours positively impacts solar PV adoption at the subregional level in both the United States and Australia \cite{graziano_spatial_2015,fuentes_solar_2024}.

Average household size has a mixed impact on solar PV deployment. In the UK, it is positively correlated with deployment at the subregional level \cite{collier_distributed_2023}. This positive relationship could be influenced by larger energy bills or increased daytime electricity usage. Conversely, a negative relationship is observed at the regional and subregional levels in the UK and at the subregional level in Australia \cite{balta-ozkan_regional_2015,alderete_peralta_spatio-temporal_2022,fuentes_solar_2024}. This negative relationship may be explained by larger households having reduced cash flow or prioritizing aesthetics over cost savings from solar PV systems \cite{balta-ozkan_regional_2015}.

Household size may also be linked to age demographics, which play a significant role in solar PV adoption. Individuals above age 40 are more likely to have greater access to cash for solar PV investments, while those aged 25–40 may face financial constraints that limit their ability to install solar PV systems. In England, a higher share of the population above age 40 is associated with greater PV installation at the subregional level \cite{laura_williams_identifying_2012}. Conversely, in Australia, a higher share of the population aged 25–40 has a negative impact on PV adoption at the subregional level \cite{fuentes_solar_2024}.

Income shows a mixed relationship with solar PV deployment. In the UK, income is positively correlated with PV adoption at the subregional and regional levels, and similarly, a positive correlation is observed at the subregional level in the United States \cite{alderete_peralta_spatio-temporal_2022,balta-ozkan_regional_2015,yu_deepsolar_2018}. However, a negative correlation between income and solar PV adoption has been found at the subregional level in Australia \cite{fuentes_solar_2024}. This mixed relationship may be explained by differences in motivation. While individuals with higher incomes can afford to install solar PV systems, they may choose not to if they are not concerned about cost savings or if they dislike the aesthetics of solar panels.

GDP is strongly correlated with solar PV capacity at the country level globally and in China \cite{alghanem_global_2024,liu_forecasting_2022}. However, in other studies, GDP does not show a proportional relationship to a country's solar PV capacity \cite{celik_review_2020}. At the regional level in Germany, GDP is negatively correlated with PV deployment \cite{mayer_deepsolar_2020}.

Industrial added value is positively correlated with solar PV deployment at the country level \cite{alghanem_global_2024,liu_forecasting_2022}. This relationship may reflect the fact that countries with higher industrial output tend to have more resources and investments available for renewable energy technologies, including solar PV. Additionally, industrial sectors often benefit from renewable energy adoption both through direct use in manufacturing processes and through the promotion of green technologies, which may further drive solar PV deployment.

Electricity consumption is positively correlated with solar PV deployment at the country level globally \cite{alghanem_global_2024}. This relationship extends to the regional and subregional levels in the UK, where higher electricity consumption is also associated with greater PV deployment \cite{alderete_peralta_spatio-temporal_2022,laura_williams_identifying_2012,balta-ozkan_regional_2015,collier_distributed_2023}.

Solar radiation has been found to have a positive correlation with solar PV deployment at the regional and subregional levels \cite{westacott_novel_2016,yu_deepsolar_2018,aklin_geography_2018,balta-ozkan_regional_2015}. This is logical, as regions with abundant sunlight provide more favourable conditions for solar power generation, making solar PV an attractive energy option. However, other studies suggest that a country's solar radiation potential is not always proportional to solar PV deployment \cite{celik_review_2020,alghanem_global_2024}. This discrepancy may be linked to the geography of a country or region. In countries with abundant agricultural land receiving high levels of solar radiation, there is often more solar PV capacity, as large open spaces are ideal for solar farms. In contrast, countries such as Austria, where two-thirds of the land is covered by the Alpine mountains \cite{kanonier_arthur_spatial_2018}, and where solar radiation is concentrated in urban areas, may face challenges in deploying solar PV, as urban environments typically offer limited space for large-scale installations. 

% In contrast, countries such as Japan and the Netherlands, where solar radiation is concentrated in urban areas may face challenges in deploying solar PV, as urban environments typically have limited space for large-scale installations.

Rural areas are positively correlated with solar PV deployment \cite{westacott_novel_2016,laura_williams_identifying_2012,aklin_geography_2018}. This is likely due to the availability of space for solar installations and the higher likelihood that rural areas have houses with suitable rooftops for PV systems. 

Solar PV is correlated with agricultural areas globally \cite{kruitwagen_global_2021,alghanem_global_2024,van_de_ven_potential_2021}, as well as with the gross value added by agriculture in Germany \cite{mayer_deepsolar_2020}. This is expected, as large-scale PV installations are often sited on agricultural land.

These studies show that the relationship between solar PV deployment and geographical factors is complex and varies depending on the geographical region and analysis resolution (national, regional, subregional). These factors are crucial in determining where and how solar PV systems are installed. By incorporating them into a comprehensive model, we can better predict regional solar PV deployment patterns, identify areas with high potential for future installations, and uncover barriers to deployment in underserved regions.

None of the existing studies comprehensively account for all types of solar PV capacity (residential, commercial, and utility-scale) installed in a country. A key challenge in modelling regional and subregional PV capacity is the limited availability of relevant social, economic, climatic, and land use data. This study introduces the first comprehensive regional PV capacity model that disaggregates national capacity across 168 NUTS 3 regions. The model will be used to allocate unknown capacity to specific geographic regions, as a benchmarking tool for assessing regional performance, and to serve as a forecasting tool.

% None of the existing studies on the UK comprehensively account for all types of solar PV capacity (residential, commercial, and utility-scale) across England, Wales, and Scotland. A key challenge in modelling regional and subregional capacity is the limited availability of high-resolution data for these regions. To address this limitation, this study uses land cover data as a proxy for geographical factors. It is the first to model regional PV capacity at the NUTS 3 level across Great Britain using publicly available datasets. The models developed will be used to disaggregate national solar PV capacity into regional levels, function as benchmarking tools for assessing regional performance, and serve as forecasting tools.

\section{Methodology}
\label{methodology}

We previously reviewed and analysed the relationship between social, economic, climatic, and land use factors and national solar PV deployment, concluding that these variables can effectively model national PV capacity \cite{alghanem_global_2024}. To identify factors associated with regional solar PV deployment in Great Britain, we adopt a similar methodology, analysing the relationships between these geographical factors and solar PV deployment at the NUTS 3 regional level in Great Britain. Given the potential non-linear relationships between these factors and PV capacity, we analyse these relationships using both Pearson and Spearman correlation coefficients. Additionally, we calculate the coefficient of determination ($R^2$) by fitting linear regression models between each feature and solar PV capacity. We select the features with the highest correlation averages and use them to train an XGBoost model that estimates the regional capacity percentage for 168 GB regions over the period from 2010 to 2023.

\subsection{Data and Data Processing}

Data for our analysis were obtained from various publicly accessible sources. Climate variables such as global horizontal irradiance (GHI), wind speed, total precipitation, sea-level pressure, and 2-meter air temperature were sourced from the Copernicus Climate Change Service \cite{copernicusCopernicusClimate}. Land use data came from the CORINE Land Cover 2018 dataset \cite{copernicusCORINELand}. Economic indicators such as regional gross value added by industry (GVA) were accessed from the UK's office of national statistics \cite{office_for_national_statistics_regional_2024}. Renewables Obligation Certificates (ROCs) buy-out prices and obligation levels were sourced from the Office of Gas and Electricity Markets (ofgem) \cite{ofgem_renewables_2024}. Solar PV capacity data were gathered from Sheffield Solar \cite{sheffield_solar_sheffield_2024}.

Our analysis is framed within the NUTS 2021 (Nomenclature of Territorial Units for Statistics) classification, a hierarchical system used to divide the European Union’s territory for statistical purposes. NUTS regions are classified into three levels, each defined by specific population thresholds to ensure statistical consistency \cite{noauthor_common_2024}. For this study, we focused on NUTS 3 regions, with populations ranging between 150,000 and 800,000 people, as data at this level are available in GB compared to a higher resolution in which case data availability becomes an issue. All data were processed according to the NUTS 3 regional classification and aggregated annually.

The CORINE land cover dataset provides a three-tier hierarchical classification system. Level 3, with its 44 thematic classes, offers detailed analysis but has a thematic accuracy of over 85\% \cite{adrien_moiret-guigand_validation_2021}, potentially leading to some misclassification. Higher levels, such as Levels 1 and 2, have fewer categories, which generally increases thematic accuracy \cite{aune-lundberg_content_2021,perez-hoyos_incorporating_2014,latifovic_accuracy_2004} and reduces misclassification risk. However, these broader classifications provide less detail on specific land use types, which is crucial for this study. To address this, we incorporate all three levels into the analysis: Level 3 provides detailed insights into specific land cover types, while Levels 1 and 2 offer a broader context to help mitigate misclassification risks.

Capacity data were processed at both the NUTS 3 (regional) and NUTS 0 (national) levels, with regional capacity at the NUTS 3 level illustrated in \Cref{fig:3a}. Gross Value Added (GVA) data were already available at the International Territorial Level (ITL), which corresponds to NUTS 3 regions \cite{office_for_national_statistics_international_nodate}. Climate data, originally available at the NUTS 2 level, were assumed to be uniform across all NUTS 3 regions within a given NUTS 2 region.

We normalize the data in our dataset either to the national total or the national average, depending on the variable. Climate data were normalized relative to national averages to highlight significant regional climatic variations. Economic factors, such as GVA, were normalized by national totals. Finally, regional solar PV capacities were expressed as a proportion of the national total capacity.

% We constructed two primary datasets: one containing absolute values and another where values were normalized to either the national total or national average, depending on the variable. This dual dataset approach facilitates a more detailed analysis by accounting for both the absolute contributions of various factors and their relative importance at a national level. The datasets cover 168 GB regions, and span the years 2010 to 2023.

% We normalize the data in our dataset to either the national total or national average, depending on the variable. climate data were normalized relative to national averages to highlight significant regional climatic variations. Economic factors, such as GVA were normalized by national totals. Finally, regional solar PV capacities were normalized relative to the national total capacity.

% In the normalized dataset, climate data were normalized relative to national averages to highlight significant regional climatic variations. Economic factors, such as GVA were normalized by national totals. Finally, regional solar PV capacities were normalized relative to the national total capacity.

\subsection{Feature Selection}

The feature selection process adhered to the DAMA data quality framework \cite{government_data_quality_hub_meet_2021}, which considers key aspects such as accuracy, completeness, uniqueness, consistency, timeliness, and validity. Features that were complete for the majority of regions and consistent over time were prioritized over those that were only available for certain regions or specific years. Selection was based on two main criteria: correlation with PV capacity and data availability. Since the relationship between certain features and PV capacity can vary — some features exhibiting strong Spearman correlations but weaker Pearson correlations, or vice versa — we calculate the average of both correlations. This average helps capture both linear and non-linear associations, and it is used to select the most relevant features. A range of correlation thresholds were tested to identify the best model. The correlation thresholds tested ranged from 0.2 to 0.5, while the data availability threshold was set at 90\%.

\subsection{Model Training}

We selected extreme gradient boosted parallel tree algorithm (XGBoost) for our modelling approach due to several key advantages. Firstly, decision tree-based models such as XGBoost are nonparametric, meaning they do not assume any specific data distribution \cite{Imam2024}. This is crucial for our study as our data is often not normally distributed, making XGBoost a better fit compared to parametric models such as linear regression, which assume a normal distribution. Secondly, XGBoost performs exceptionally well with tabular data \cite{xgboost}, which matches the structure of our dataset.

Data were grouped by year, with the period from 2010 to 2020 used for training and 2021 to 2023 reserved for testing. An XGBoost model was trained, with hyperparameters fine-tuned using grid search and 10-fold cross-validation to optimize performance. The trained model predicts the percentage of solar PV capacity within a NUTS 3 region and is referred to as the "unscaled model". To ensure regional predictions sum to the national capacity, we scaled the predicted regional capacities so that their total matched the national value. This approach is referred to as the "scaled model".

% The dataset was used to develop a model for predicting the percentage of solar PV capacity within a NUTS 3 region. Data were grouped by year, with the period from 2010 to 2020 used for training and 2021 to 2023 reserved for testing. An XGBoost model was trained on this dataset, with its hyperparameters fine-tuned using grid search and 10-fold cross-validation to optimize performance. We refer to this as the "unscaled model". To ensure that the model’s regional predictions sum to the national capacity, we scaled the predicted regional capacities so that their total matched the national value. This approach is referred to as the "scaled model".

% The normalized dataset was used to develop a model to predict the percentage of solar PV capacity within a NUTS 3 region. Data were grouped by year, with the period from 2010 to 2020 allocated for training and 2021 to 2023 reserved for testing. An XGBoost model was trained on this dataset, and its hyperparameters were fine-tuned using grid search with 10-fold cross-validation to optimize performance. To ensure that the model's regional predictions sum to the national capacity, we scaled the predicted regional capacities so that their sum equals the national total. This approach is referred to as the "scaled model".

We used Shapley Additive Explanations (SHAP) values to measure feature importance. SHAP is a game-theoretic approach designed to explain the output of any machine learning model, where the SHAP values are additive, meaning the contribution of each feature sums to the difference between the model's prediction and the average prediction \cite{lundberg_unified_2017}. 

We used principal component analysis (PCA) to analyse selected features and assess whether they form distinct clusters. If clustering is observed, we were able to further examine the nature and structure of these clusters to help explain the model performance.

To evaluate the model's performance, we used several error metrics, including Root Mean Squared Error (RMSE), Mean Squared Error (MSE), Mean Absolute Error (MAE), and Mean Absolute Percentage Error (MAPE). We avoid using MAPE when evaluating regional capacity percentage because it can yield large values due to data representation in percentages. Instead, MAPE is specifically applied to evaluate national capacity. 

\subsection{Model Applications}

The models can be applied in several ways: to disaggregate national capacity into regional capacity, to allocate capacity with an unknown location to a geographical region, and to serve as a benchmarking tool. A Solar PV Deployment Index (SPVDI) was developed previously to assess national capacity in a global context \cite{alghanem_global_2024}. In this study, we adapt the same concept to evaluate regional PV capacity within GB. 

The SPVDI serves as a benchmarking tool by comparing solar PV deployment in a region relative to other regions with similar social, economic, climatic, and land use characteristics. It is calculated as shown in \Cref{eq:SPDVI}, where $t_{\text{1}}$ represents the initial year and $t_{\text{2}}$ the final year of analysis. The resulting index value indicates whether a region's capacity is above or below expectations, with a positive value indicating higher-than-expected capacity and a negative value indicating lower-than-expected capacity. The SPVDI allows for performance comparisons of regions over multiple years and time periods and serves as a tool to rank regions based on their solar PV deployment.

\begin{equation}
    SPVDI = \sum_{t_{\text{1}}}^{t_{\text{2}}} (\text{{Actual Capacity - Predicted Capacity}} )_{per\ region}
    \label{eq:SPDVI}
\end{equation}

\section{Results and Discussion}
\label{results}

\Cref{tab:norm_data} presents the analysis results for the normalized regional data, with features ranked from highest to lowest based on their average correlation. Gross value added by veterinary activities emerges as the most strongly correlated feature. Additionally, regions with higher percentages of agricultural areas and artificial surfaces tend to have a greater share of PV capacity.

Interestingly, climatic factors such as GHI ranks 18, while air temperature ranks 24 in terms of average correlation. The Spearman correlation for GHI and temperature data is higher than the Pearson correlation, suggesting a non-linear relationship between climatic factors and capacity. For GHI, the Pearson correlation is 0.37, while the Spearman correlation is 0.51. Similarly, for air temperature, the Pearson correlation is 0.37, and the Spearman correlation is 0.47.

The reason climatic factors rank lower than other factors may be attributed to threshold effects. Regions need to meet certain levels of solar resource potential to justify investments in PV systems. Once these thresholds are reached, other factors - such as land availability or socioeconomic characteristics - become more critical in determining the proportional share of PV deployment. Previous studies have identified a radiation threshold of 4.5 kWh/m²/day above which deployment is triggered \cite{yu_deepsolar_2018}.

The strong correlation between artificial surfaces and solar PV capacity can be attributed to the inclusion of energy production and distribution infrastructure within this category \cite{european_environment_agency_corine_nodate}. These features could act as proxies for grid connection points, which are crucial for integrating grid-connected solar PV systems. Regions with more access to such infrastructure are better positioned to support large-scale solar installations. Furthermore, the presence of industrial agricultural facilities, which are also classified as artificial surfaces, likely contributes to the observed correlation with PV capacity. The correlation between agricultural land and solar PV capacity is strong. This is expected since commercial, industrial, and utility-scale solar PV projects are predominantly situated on agricultural land, particularly arable land \cite{van_de_ven_potential_2021,kruitwagen_global_2021}.

Gross value added by veterinary activities is correlated with artificial surfaces and agricultural land, which might explain why it ranks the highest in terms of correlation with PV capacity. This could also be linked to the presence of farm animals, as veterinary activities are more likely to be concentrated in regions with higher demand, such as areas with farms and extensive agricultural land. 

The features that were selected for the model are: gross value added by veterinary activities, arable land (21), non-irrigated arable land (211), urban fabric (11), artificial surfaces (1), discontinuous urban fabric (112), agricultural Areas (2), sport and leisure facilities (142), artificial, non-agricultural vegetated areas (14), and mine, dump and construction sites (13).

% The features that were selected for the model are: gross value added by veterinary activities, Arable land (21), Non-irrigated arable land (211), Urban fabric (11), artificial surfaces (1), Discontinuous urban fabric (112), Agricultural Areas (2), Sport and leisure facilities (142), Artificial, non-agricultural vegetated areas (14), and mine, dump and construction sites (13).

\Cref{tab:3} shows the regional error metrics for the unscaled and scaled models. Scaling doesn't improve the model significantly. However, we still use it in cases where the national capacity is being disaggregated. The model explains 89\% of the variation in NUTS 3 regional PV capacity, has a mean average error of about 20 MW, and an root mean square error of 41 MW. 

% \Cref{tab:3} shows the regional error metrics for the normalized model. Scaling doesn't improve the model significantly. However, we still use it in cases where the national capacity is being disaggregated. The model explains 89\% of the variation in NUTS 3 regional PV capacity, has a mean average error of about 20 MW, and an root mean square error of 41 MW. 

\begin{table}[] 
% \caption{Regional error metrics for different models. The Normalized Model metrics are reported in both megawatts (MW) and percentages (\%). For the Scaled models, regional predictions are adjusted to ensure they sum to the national capacity. Metrics include: Training \( R^2 \), Test \( R^2 \), Mean Absolute Error (MAE), Mean Squared Error (MSE), and Root Mean Squared Error (RMSE).}

\caption{Regional error metrics for both the unscaled and scaled models. The metrics are reported in both megawatts (MW) and percentages (\%). For the Scaled models, regional predictions are adjusted to ensure they sum to the national capacity. Metrics include: Training \( R^2 \), Test \( R^2 \), Mean Absolute Error (MAE), Mean Squared Error (MSE), and Root Mean Squared Error (RMSE).}

\label{tab:3}
\resizebox{\textwidth}{!}{%
\begin{tabular}{lllllll}
\hline
Metric & Training $R^2$ & Test $R^2$ & MAE & MSE & RMSE \\ \hline
Unscaled Model (MW) &
   0.99 &
   0.89 &
   20.27&
   1701 &
  41.25
   \\
Scaled Model (MW) &  0.99   &  0.90  & 19.94 &1610 & 40.13 \\
Unscaled Model (\%) &
   0.98 &
   0.90 &
   0.13 &
   0.07 &
  0.26
   \\
Scaled Model  (\%)&   0.98  &  0.91  &  0.13 & 0.06  & 0.25 \\ \hline

\end{tabular}%
}
\end{table}

\begin{figure}[]
    \centering
    
    \subfloat[Unscaled model.]{
        \includegraphics[width=0.49\linewidth]{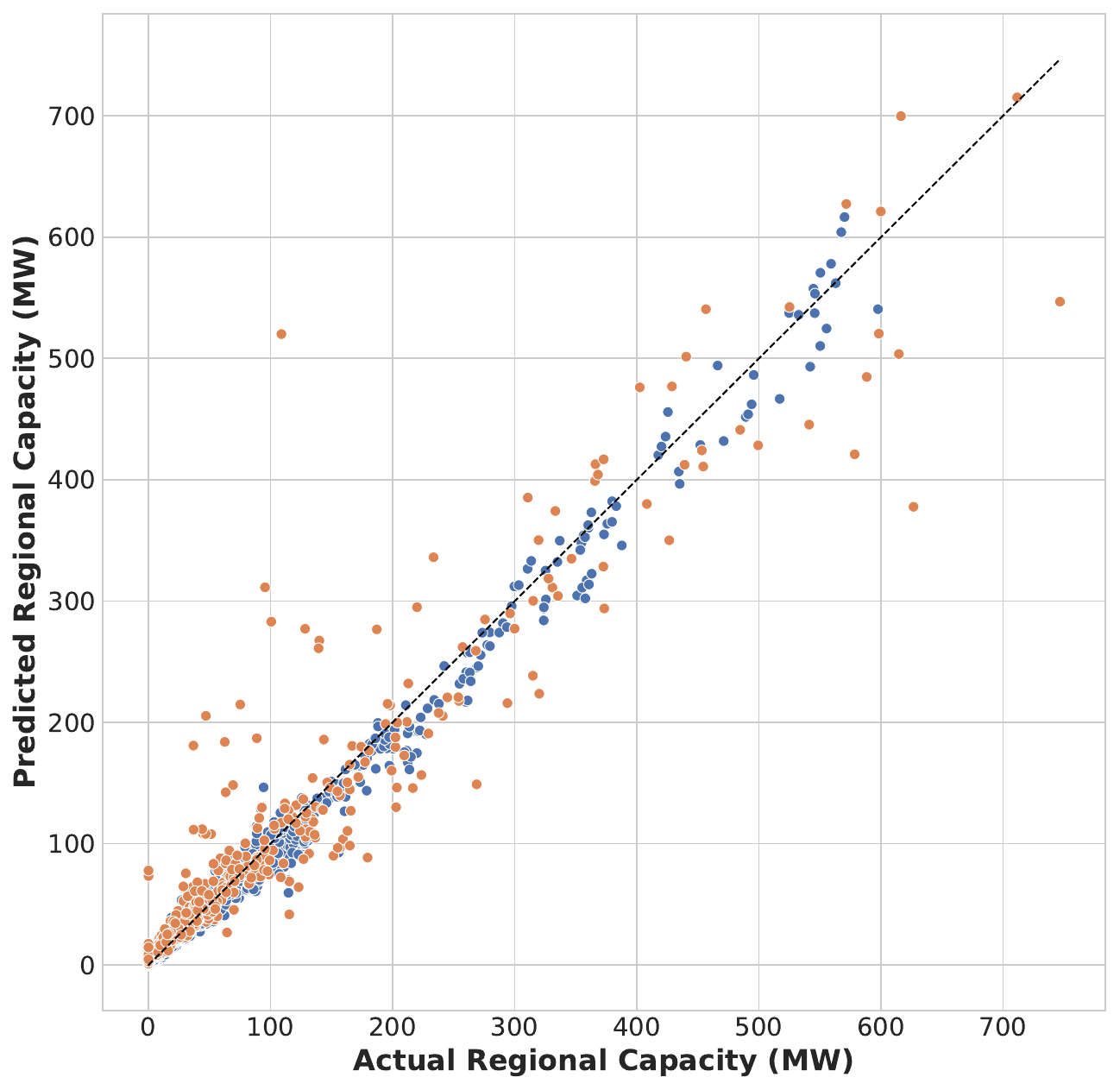}
        \label{fig:1a}
    }
    % \hfill
    \subfloat[Scaled model.]{
        \includegraphics[width=0.49\linewidth]{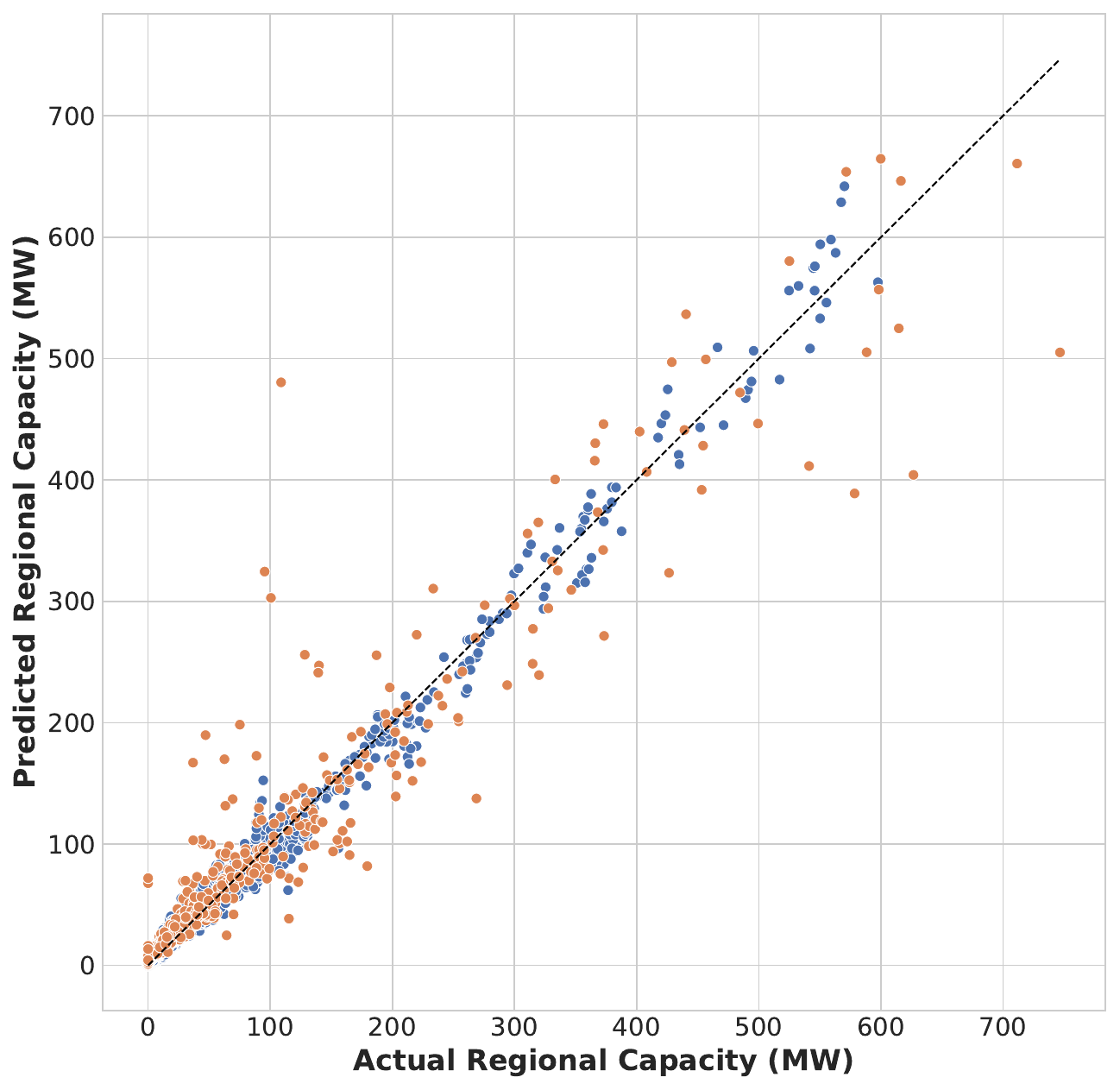}
        \label{fig:1b}
    }
    
    \caption{Actual versus predicted regional photovoltaic capacity for 168 NUTS 3 regions in Great Britain, covering the years 2010 to 2023. Blue points represent training data, while orange points represent test data.}
    \label{fig:1}
\end{figure}

\Cref{fig:1} shows the actual vs estimated regional capacity for the unscaled and scaled models. The unscaled model performs well and scaling does not make any significant changes to the the regional estimates.  

% \Cref{fig:1} shows the actual vs estimated regional capacity for the normalized model before and after scaling. The model performs well and scaling does not make any significant changes to the the regional estimates.  

\Cref{tab:4} shows the national error metrics for the unscaled and scaled models. The unscaled model has an error of 5.4\% in estimating national capacity. Interestingly, this is very similar to the 5\% error reported by \citet{huxley_uncertainties_2022}. 

% \Cref{tab:4} shows the national error metrics for the normalized model. The normalized model has an error of 5.4\% in estimating national capacity. Interestingly, this is very similar to the 5\% error reported by \citet{huxley_uncertainties_2022}. 

\begin{table}[]
\caption{National error metrics for the unscaled and scaled models. These metrics compare the actual national capacity with the estimated national capacity, which is derived from summing the regional capacities. Scaling refers to adjusting regional capacities so that their sum matches the actual national capacity.}

% \caption{National error metrics. These metrics compare the actual national capacity with the estimated national capacity, which is derived from summing the regional capacities. Scaling refers to adjusting regional capacities so that their sum matches the actual national capacity.}

\label{tab:4}
\resizebox{\textwidth}{!}{
\begin{tabular}{lllllll}
\hline
Metric                         & R$^2$   & MAE (MW) & MSE (MW\(^{2}\)) & RMSE (MW) & MAPE (\%) \\ \hline
Unscaled Model (entire dataset) & 0.99 &  441   &  322000  &   568   & 5.4  \\
Scaled Model (entire dataset) & 1 & 0 & 0 & 0 & 0.0 \\ \hline
\end{tabular}
}
\end{table}

The feature importance analysis reveals that non-irrigated arable land (211) accounts for 36\% of the total SHAP values, followed by discontinuous urban fabric (112) at 21\%, artificial surfaces (1) at 12\%, sport and leisure facilities (142) at 7\%, agricultural areas (2) at 7\%, gross value added by veterinary activities at 6\%, urban fabric (11) at 5\%, mine, dump, and construction sites (13) at 4\%, non-agricultural vegetated areas (14) at 2\%, and arable land (21) at 0\%. Furthermore, the analysis highlights that artificial surfaces have the most significant overall impact, accounting for 51\% of the total SHAP values. This is followed by agricultural land, which contributes 43\%, and gross value added by veterinary services, contributing 6\%.

When performing PCA and clustering the selected features, two distinct groups emerge. The first cluster includes artificial surfaces (1), discontinuous urban fabric (112), urban fabric (11), non-agricultural vegetated areas (14), sport and leisure facilities (142), and gross value added by veterinary services. The second cluster includes arable land (21), non-irrigated arable land (211), agricultural areas (2), and mine, dump, and construction sites (13). The most representative variable in the first cluster is artificial surfaces (1), which explains 96\% of the variation within the cluster. For the second cluster, the most representative variable is arable land (21), which explains 92\% of the variation in the cluster. These findings align with the results above, which highlight artificial surfaces and agricultural land as the most significant factors influencing PV deployment at the NUTS 3 level.

\subsection{Model Applications}
\begin{figure}[htbp]
    \centering
    \subfloat[]{%
        \includegraphics[width=0.3\linewidth]{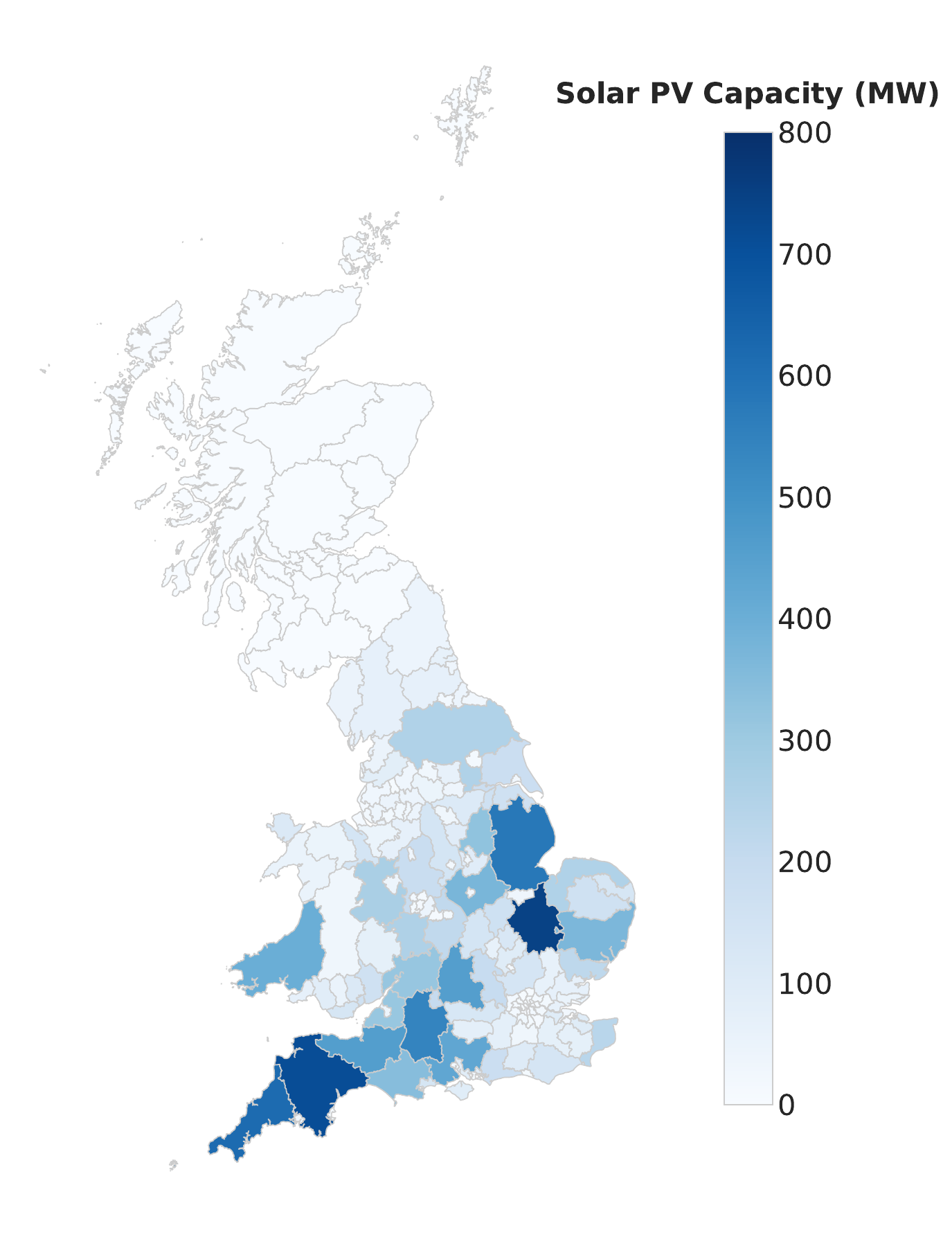}
        \label{fig:3a}
    }
    % \hfill
    \subfloat[]{%
        \includegraphics[width=0.3\linewidth]{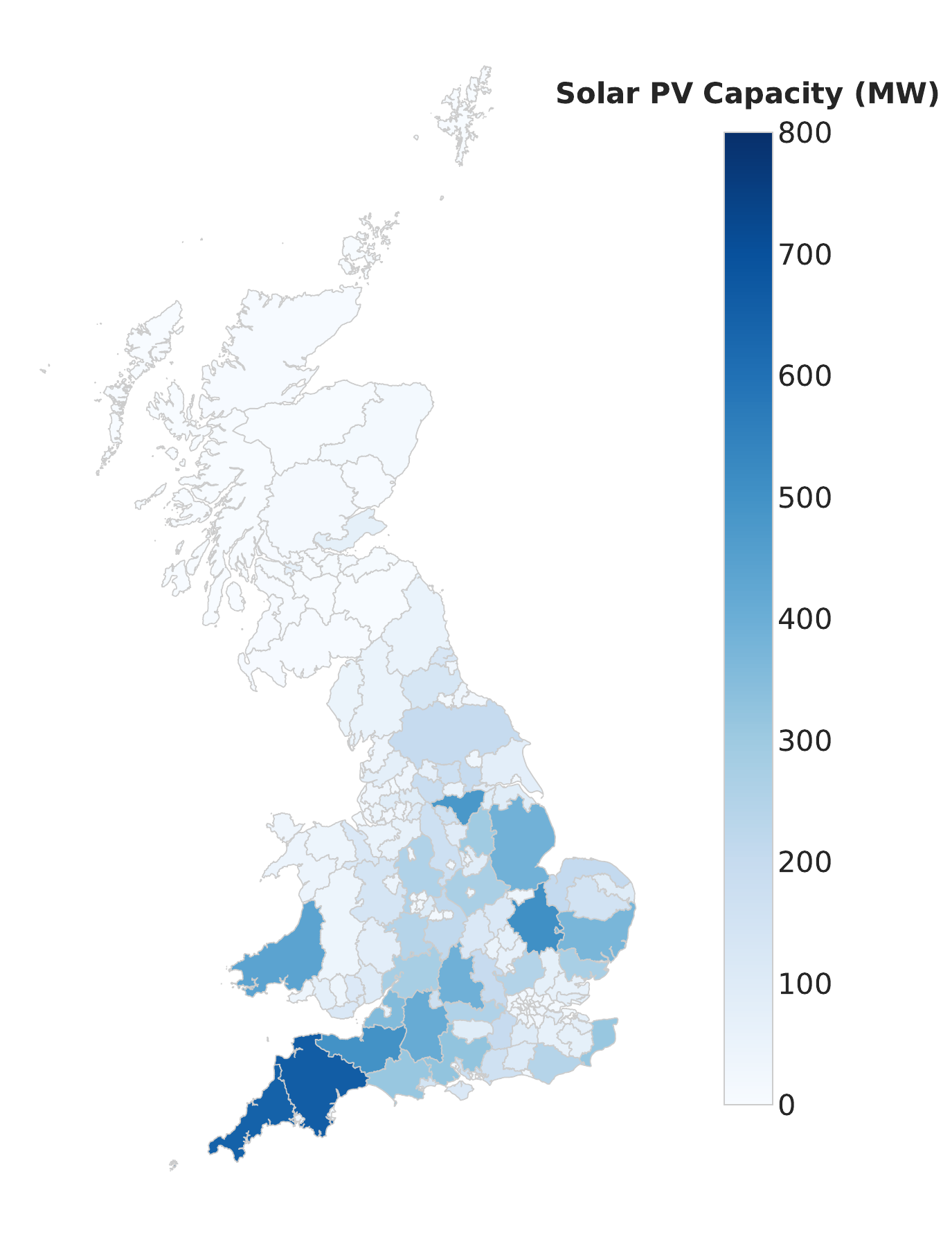}
        \label{fig:3b}
    }
    % \hfill
    \subfloat[]{%
        \includegraphics[width=0.275\linewidth]{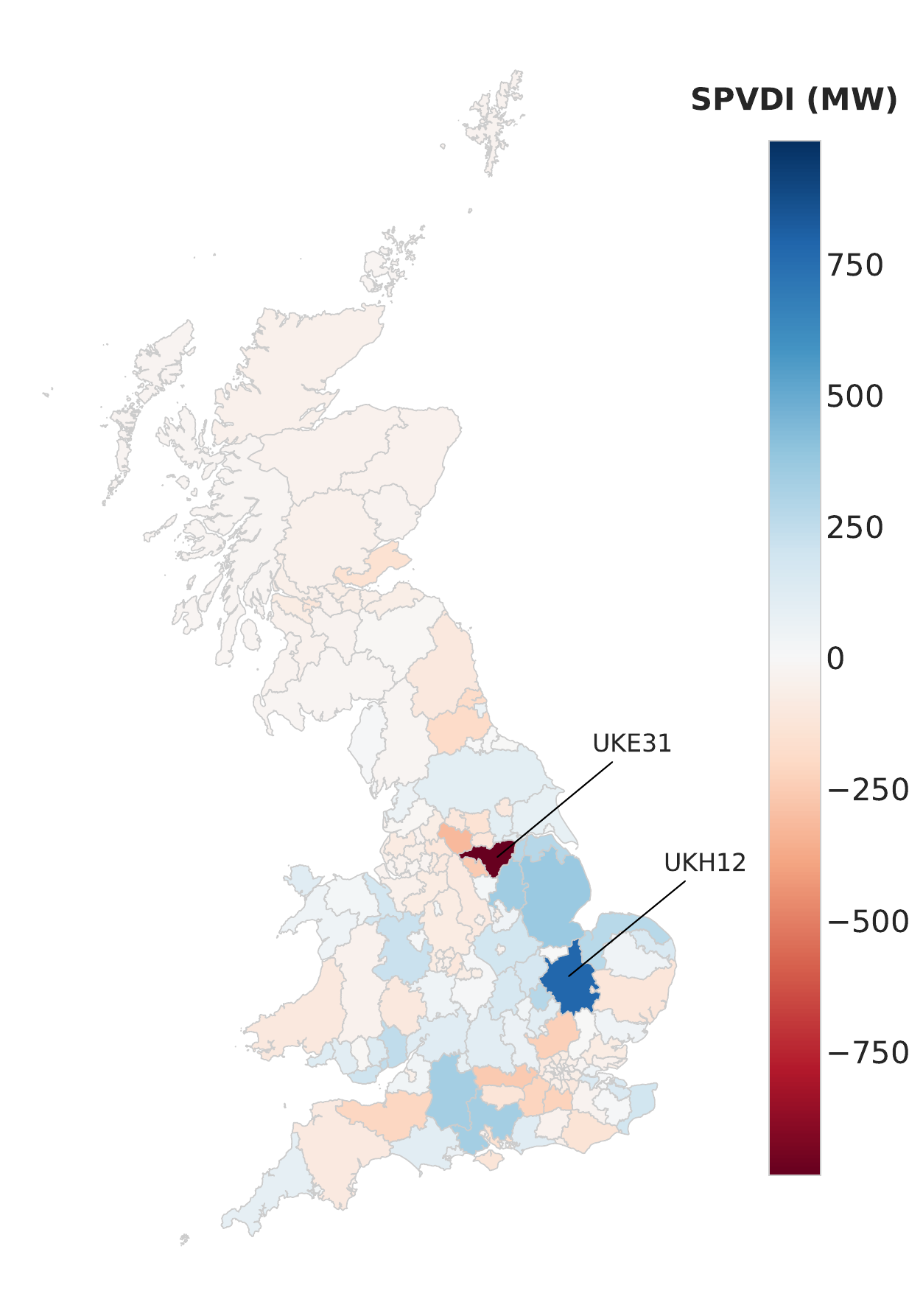}
        \label{fig:3c}
    }
    \caption{(a) Actual GB regional solar photovoltaic capacity for 168 NUTS 3 regions in 2023 (b) Predicted GB regional solar photovoltaic capacity for 168 NUTS 3 regions in 2023 (c) Solar PV deployment index results for the years 2010 to 2023. Positive values indicate regions where the actual capacity exceeds the expected capacity, while negative values indicate regions where the actual capacity is less than expected. The regions included are Barnsley, Doncaster and Rotherham (UKE31) which under deployed by 985 MW and Cambridgeshire CC (UKH12) which over deployed by 780 MW.}
    \label{fig:3}
\end{figure}

\begin{figure}[]
    \centering
    
        \includegraphics[width=0.5\linewidth]{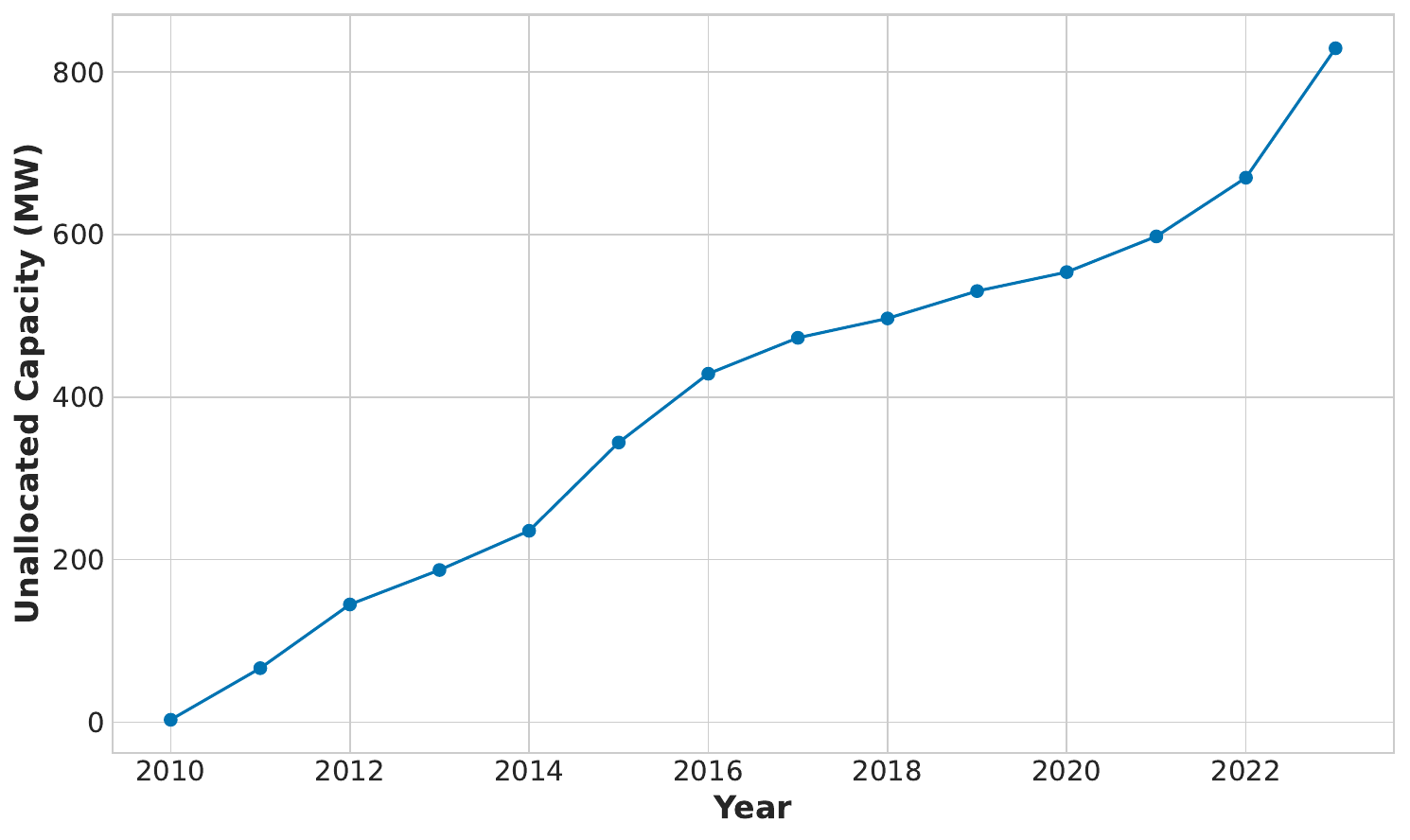}
        \label{fig:4a}

    \caption{Unallocated solar photovoltaic capacity for 168 NUTS 3 regions in Great Britain from the year 2010 to 2023. }
    \label{fig:4}
\end{figure}

\Cref{fig:3a} illustrates the actual regional capacities for the regions used in training and testing the model for 2023. In some cases, the locations of solar PV systems are not recorded. \Cref{fig:4} shows the unallocated capacity per year from 2010 to 2023. Initially, this wasn't an issue due to the low capacity, but as it grows, it becomes more difficult to monitor generation accurately. As PV capacity increases, it becomes increasingly important to know the geographical distribution of these capacities due to their impact on grid stability. For example, in 2023, these unrecorded systems accounted for 829 MW in GB. The scaled model was used to allocate these capacities, leading to updated regional capacity estimates, as depicted in \Cref{fig:3b}. We provide a dataset of regional PV capacities for 168 GB regions spanning from 2010 to 2023, enabling more accurate generation monitoring at the regional level. Unlike simple allocation methods that distribute PV capacity based on the percentage of deployment per region, our model accounts for the unique geographical factors that affect capacity diffusion. This approach addresses systematic issues, such as inconsistent reporting, under-reporting in less monitored regions, and challenges in capturing small-scale installations, ensuring more reliable and precise regional insights.

% \Cref{fig:3a} illustrates the actual regional capacities for the regions used in training and testing the model for 2023. In some cases, the locations of solar PV systems are not recorded. \Cref{fig:4} shows the unallocated capacity per year from 2010 to 2023. Initially, this wasn't an issue due to the low capacity, but as it grows, it becomes more difficult to monitor generation accurately. As PV capacity increases, it becomes increasingly important to know the geographical distribution of these capacities due to their impact on grid stability. For example, in 2023, these unrecorded systems accounted for 829 MW in GB. The scaled model was used to allocate these capacities, leading to updated regional capacity estimates, as depicted in \Cref{fig:3b}. We provide a dataset of regional PV capacities for 168 GB regions spanning from 2010 to 2023, enabling more accurate generation monitoring at the regional level. Unlike simple allocation methods that distribute PV capacity based on the percentage of deployment per region, our model accounts for the unique geographical factors that affect capacity reporting. This approach addresses systematic issues, such as inconsistent reporting, under-reporting in less monitored regions, and challenges in capturing small-scale installations, ensuring more reliable and precise regional insights.

% The model serves as a valuable benchmarking tool to assess regional solar PV deployment, allowing for the evaluation of areas that exceed or fall short of expected capacities. \Cref{fig:3c} illustrates the solar PV deployment index for the period between 2010 and 2023, highlighting the top and bottom regions based on capacity differences. The top ten performing regions, exceeding expected capacity by significant margins, include Cambridgeshire CC (UKH12) with 780 MW, Lincolnshire (UKF30) with 368 MW, and North Nottinghamshire (UKF15) with 350 MW, Central Hampshire (UKJ36) with 332 MW, Wiltshire CC (UKK15) with 332 MW, and Bedford (UKH24) with 283 MW, North and North East Lincolnshire (UKE13) with 278 MW, North and West Norfolk (UKH16) with 272 MW, Monmouthshire and Newport (UKL21) with 252 MW, and Shropshire CC (UKG22) with 223 MW. In contrast, the bottom ten regions, which fall short of expected capacity, include Barnsley, Doncaster, and Rotherham (UKE31) with the largest deficit at -985 MW. Other underperforming regions are Calderdale and Kirklees (UKE44) with -316 MW, Berkshire (UKJ11) with -260 MW, Sheffield (UKE32) with -255 MW, Hertfordshire (UKH23) with -233 MW, East Surrey (UKJ26) with -227 MW, West Surrey (UKJ25) with -218 MW, Somerset (UKK23) with -209 MW, Tyneside (UKC22) with -190 MW, and Durham CC (UKC14) with -189 MW. 

The unscaled model serves as a valuable benchmarking tool to assess regional solar PV deployment, allowing for the evaluation of areas that exceed or fall short of expected capacities. \Cref{fig:3c} illustrates the solar PV deployment index for the period between 2010 and 2023, highlighting the top and bottom regions based on capacity differences. The top ten performing regions, exceeding expected capacity by significant margins, include Cambridgeshire CC (UKH12) with 780 MW, Lincolnshire (UKF30) with 368 MW, and North Nottinghamshire (UKF15) with 350 MW, Central Hampshire (UKJ36) with 332 MW, Wiltshire CC (UKK15) with 332 MW, and Bedford (UKH24) with 283 MW, North and North East Lincolnshire (UKE13) with 278 MW, North and West Norfolk (UKH16) with 272 MW, Monmouthshire and Newport (UKL21) with 252 MW, and Shropshire CC (UKG22) with 223 MW. In contrast, the bottom ten regions, which fall short of expected capacity, include Barnsley, Doncaster, and Rotherham (UKE31) with the largest deficit at -985 MW. Other underperforming regions are Calderdale and Kirklees (UKE44) with -316 MW, Berkshire (UKJ11) with -260 MW, Sheffield (UKE32) with -255 MW, Hertfordshire (UKH23) with -233 MW, East Surrey (UKJ26) with -227 MW, West Surrey (UKJ25) with -218 MW, Somerset (UKK23) with -209 MW, Tyneside (UKC22) with -190 MW, and Durham CC (UKC14) with -189 MW. 

These findings align with those of \citet{collier_distributed_2023}, who modelled small-scale domestic solar PV deployment in England and Wales at the LSOA level. \citet{collier_distributed_2023} observed that local authority districts within Cambridgeshire CC (UKH12), Lincolnshire (UKF30), North Nottinghamshire (UKF15), North and North East Lincolnshire (UKE13), Monmouthshire and Newport (UKL21), Peterborough (UKH11), Essex Haven Gateway (UKH34), Breckland and South Norfolk (UKH17), Cornwall and Isles of Scilly (UKK30), Sunderland (UKC23), and Bath and North East Somerset, North Somerset, and South Gloucestershire (UKK12) exhibited higher than expected capacities. However, discrepancies were observed between our findings and those of \citet{collier_distributed_2023} for certain regions. Specifically, while their study identified local authority districts within Devon (UKK43) and Suffolk CC (UKH14) as having higher than expected capacities, our our results suggest that these regions fall short by -94.9 MW and -119.5 MW, respectively. This discrepancy may be attributed to differences in data sources, as \citet{collier_distributed_2023} focuses on residential solar PV systems with capacities of up to 10 kW, whereas our analysis includes all system types. It is possible that these areas have higher than expected capacity for residential systems but lower than expected capacity for commercial and utility-scale systems, leading to the observed differences in overall results. Both regions have a very high percentage of artificial surfaces and agricultural land compared to other regions. This high proportion might lead the model to estimate a higher capacity than is present in reality, potentially contributing to the observed shortfall.

% \Cref{fig:7} shows the SHAP analysis of the average contribution of the features from 2010 to 2023 in Cambridgeshire, and Barnsley, Doncaster, and Rotherham . 
% The model's average prediction is 0.565\% of national capacity, as determined by SHAP analysis. In Cambridgeshire, the SHAP analysis of feature contributions from 2010 to 2023 reveals that non-irrigated arable land (211) is the largest contributor to the prediction, with a mean contribution of 1.39\%. Other contributors include discontinuous urban fabric (112) at 0.74\%, sport and leisure facilities (142) at 0.44\%, artificial surfaces (1) at 0.16\%, gross value added by veterinary services at 0.11\%, construction sites (13) at -0.06\%, urban fabric (11) at 0.03\%, agricultural areas (2) at 0.02\%, non-agricultural vegetated areas (14) at -0.01\%, and arable land (21) at 0\%. This results in an average predicted capacity for Cambridgeshire of 3.39\% of the national total, equivalent to 341 MW. The high prediction for Cambridgeshire is primarily driven by non-irrigated arable land (211), but other factors, such as access to grid connection points, could also explain why this region has more capacity than expected. We hypothesize that the region's southern location, the prevalence of agricultural areas, and its proximity to transmission lines - where fewer constraints exist compared to other regions - may play a significant role. Additionally, social effects, such as peer influence, could help explain the higher than expected capacity in this area.

\begin{figure}[]
    \centering
    
    \subfloat[UKH12]{
        \includegraphics[width=0.5\linewidth]{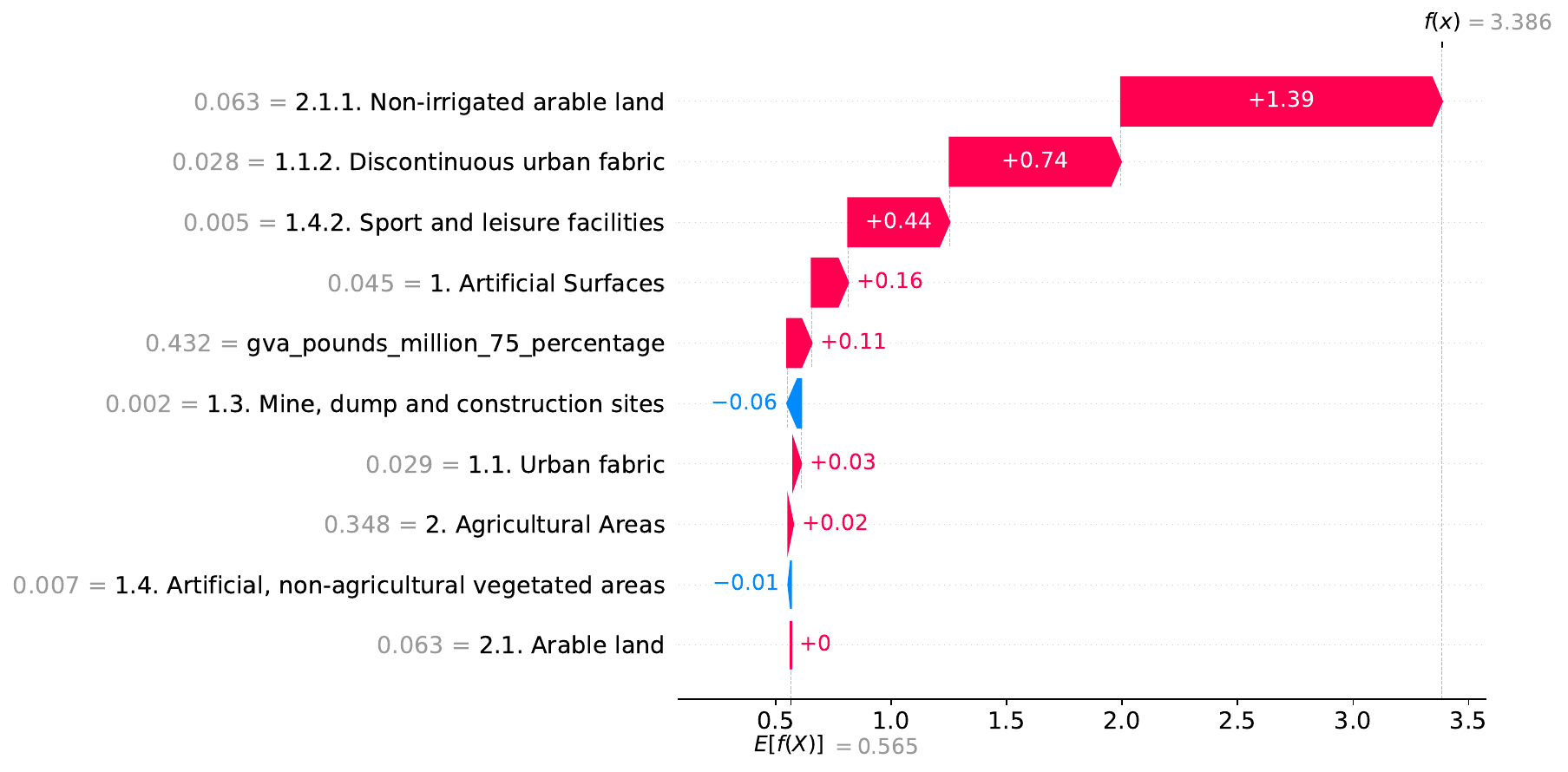}
        \label{fig:7a}
    }
    % \hfill
    \subfloat[UKE31]{
        \includegraphics[width=0.5\linewidth]{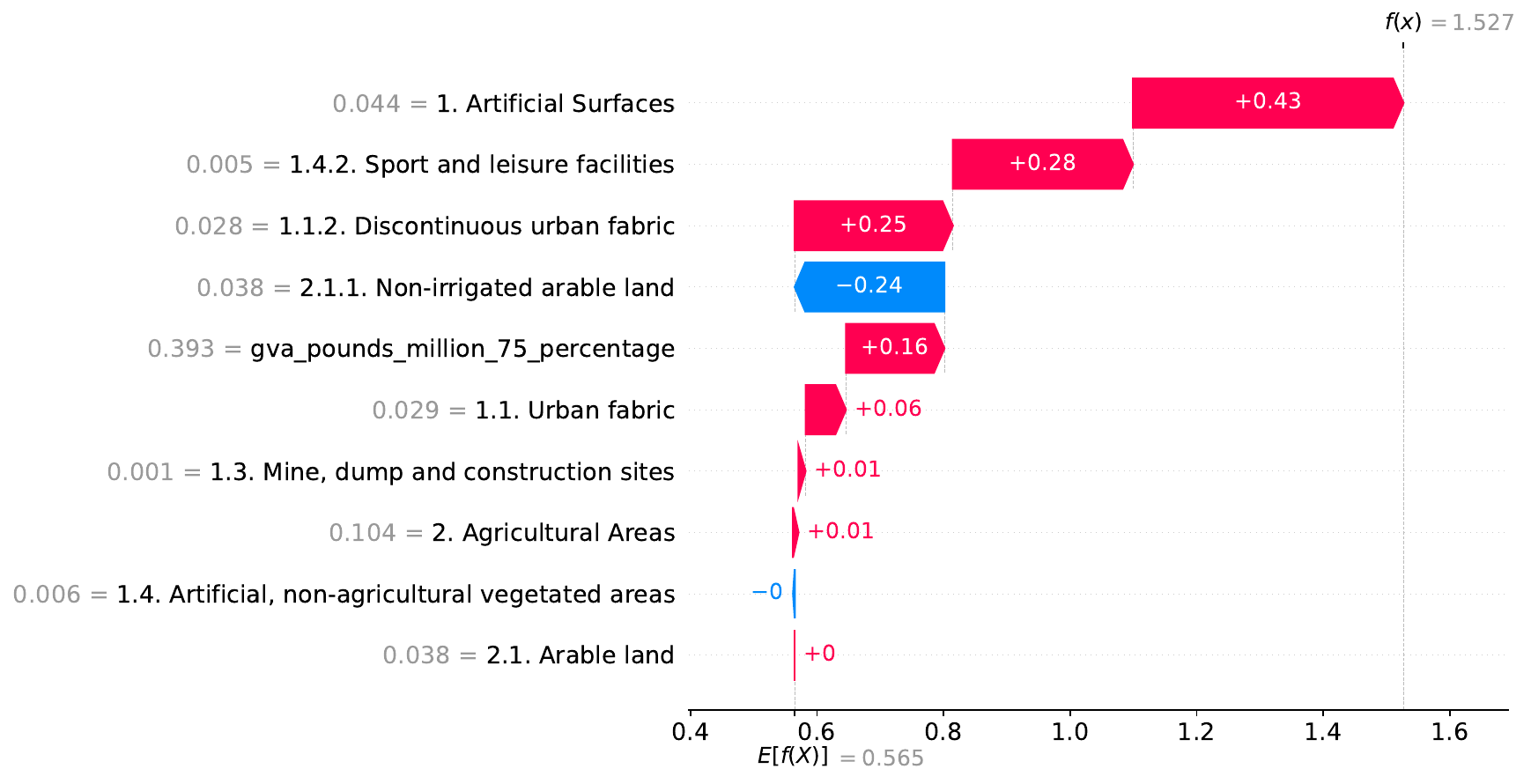}
        \label{fig:7b}
    }
    
    \caption{SHAP analysis showing the mean contribution of factors to solar PV capacity percentage between 2010 and 2023 for (a) Cambridgeshire CC (UKH12) and (b) Barnsley, Doncaster, and Rotherham (UKE31). In the SHAP framework, $E[f(x)]$ represents the model’s average prediction, while $f(x)$ is the predicted value, calculated as the sum of the contributions of the features and the model’s average prediction. }
    \label{fig:7}
\end{figure}

\Cref{fig:7} shows the SHAP analysis of the average contribution of the features from 2010 to 2023 in Cambridgeshire, and Barnsley, Doncaster, and Rotherham . 
The unscaled model's average prediction is 0.565\% of national capacity, as determined by SHAP analysis. In Cambridgeshire, the SHAP analysis of feature contributions from 2010 to 2023 reveals that non-irrigated arable land (211) is the largest contributor to the prediction, with a mean contribution of 1.39\%. Other contributors include discontinuous urban fabric (112) at 0.74\%, sport and leisure facilities (142) at 0.44\%, artificial surfaces (1) at 0.16\%, gross value added by veterinary services at 0.11\%, construction sites (13) at -0.06\%, urban fabric (11) at 0.03\%, agricultural areas (2) at 0.02\%, non-agricultural vegetated areas (14) at -0.01\%, and arable land (21) at 0\%. This results in an average predicted capacity for Cambridgeshire of 3.39\% of the national total, equivalent to an average capacity of 341 MW between 2010 and 2023. The high prediction for Cambridgeshire is primarily driven by non-irrigated arable land (211), but other factors, such as access to grid connection points, could also explain why this region has more capacity than expected. We hypothesize that the region's southern location, the prevalence of agricultural areas, and its proximity to transmission lines - where fewer constraints exist compared to other regions - may play a significant role. Additionally, social effects, such as peer influence, could help explain the higher than expected capacity in this area.

Based on SHAP analysis of the mean contribution of features between 2010 and 2023 in Barnsley, Doncaster, and Rotherham, the largest contributor is artificial surfaces (1) with a contribution of 0.43\%. This is followed by sport and leisure facilities (142) at 0.28\%, discontinuous urban fabric (112) at 0.25\%, and non-irrigated arable land (211) at -0.24\%. Other contributors include gross value added by veterinary services at 0.16\%, urban fabric (11) at 0.06\%, construction sites (13) at 0.01\%, agricultural areas (2) at 0.01\%, non-agricultural vegetated areas (14) at 0\%, and arable land (21) at 0\%. This results in an average prediction for Barnsley, Doncaster, and Rotherham of 1.53\% of national capacity, equivalent to an average capacity of 141 MW between 2010 and 2023. 

The SHAP analysis indicates that the high prediction for the region is primarily driven by artificial surfaces. The predominance of artificial surfaces in this region suggests that solar PV systems are primarily residential and commercial. This observation aligns with findings by \citet{westacott_novel_2016}, which identify domestic installations as the dominant source of capacity in the area. However, the region has 958 MW less capacity than expected, suggesting that domestic systems alone cannot explain the shortfall. Since agricultural areas are excluded based on SHAP analysis, the gap is likely attributable to commercial systems. 

To explain the shortfall, there may be unique characteristics of artificial surfaces in this region that differentiate it from others. For example, variations in business ownership or type could play a role. \citet{westacott_novel_2016} highlights that the non-domestic rooftop market is significantly underdeveloped compared to the domestic and ground-mounted markets in the UK. Additionally, grid connection constraints for commercial installations and differences in the planning permission process may further influence deployment in this region.

Finally, Barnsley, Doncaster, and Rotherham's gross disposable household income (GDHI) per head, measured as indices where the UK equals 100, ranged from 79.6 to 81.3 between 2010 and 2022 \cite{office_for_national_statistics_regional_itl_2024}. This consistently lower GDHI compared to the national average may help explain the lower than expected solar PV capacity, as reduced disposable income could limit investment in domestic installations. GDHI is excluded from the model because the data is typically published with a delay of about two years, meaning data for 2021 was only made available in September 2023. \newline

The models serve three main purposes. First, they enable the disaggregation of national PV capacity, with the scaled model being the most appropriate for this task. This is particularly useful when there is a significant amount of unallocated capacity at the national level. This can help with more accurate generation monitoring. Second, the models facilitate the benchmarking of regional capacities within GB using the unscaled model. This approach allows for ranking regions based on their PV deployment, helping to identify areas that may require interventions, such as improving access to grid connection points or policy adjustments. Finally, the scaled model can be used for forecasting capacity. This is beneficial for strategic grid expansion planning and can help in addressing land use conflicts by pinpointing regions where capacity is likely to be concentrated.

\section{Conclusion}
\label{conclusion}

This study provides a comprehensive analysis of regional solar PV capacity in Great Britain, identifying artificial surfaces and agricultural areas as the most significant factors influencing deployment at the NUTS 3 regional level. To address the need for detailed capacity estimates, we developed a model that explains 89\% of the variation in PV capacity, with a mean absolute error (MAE) of 20 MW and a national mean absolute percentage error (MAPE) of 5.4\%. Additionally, we provide a dataset of regional capacity estimates for NUTS 3 regions in Great Britain, covering the years 2010 to 2023.

The model serves three primary purposes: disaggregating national PV capacity into regional capacity, benchmarking regional solar PV deployment, and forecasting future solar PV deployment. It offers practical applications for grid operators by improving generation monitoring, which requires precise knowledge of the geographical distribution of capacity. The model also supports long-term grid network planning by identifying where capacity is likely to be distributed and enabling targeted grid expansions in these areas. Furthermore, it can help pinpoint regions where deployment falls short of expectations, offering insights into where interventions, such as policy adjustments or infrastructure investments, may be needed.

Future research could explore regions where solar PV capacity significantly exceeds or falls below expectations. These case studies could provide valuable insights into barriers to deployment and identify historical incentives or policies that have successfully driven higher deployment.

Overall, this study offers valuable tools and data to enhance operational and strategic decision-making for Great Britain's solar PV sector and lays a foundation for further research into regional capacity estimation and renewable energy planning.

\begin{landscape}
\footnotesize

\begin{longtable}{L{9cm}L{1.6cm}L{1.6cm}L{1.6cm}L{1.6cm}L{1.6cm}L{1.6cm}}
\caption{Normalized data analysis. Features considered for modelling the percentage of regional solar photovoltaic capacity. All features presented are normalized relative to national values and expressed as percentages. The data availability, coefficient of determination ($R^2$), Pearson correlation, Spearman correlation, correlation average are shown. Relevant literature that explores similar features is cited.}
\label{tab:norm_data}\\
\toprule
Feature & Data Availability (\%) & R-squared & Pearson Correlation & Spearman Correlation & Correlation Average & Relevant Literature \\
\midrule
\endfirsthead
\toprule
Feature & Data Availability (\%) & R-squared & Pearson Correlation & Spearman Correlation & Correlation Average & Relevant Literature \\
\midrule
\endhead
\midrule
% \multicolumn{7}{r}{Continued on next page} \\
% \midrule
\endfoot
\bottomrule
\endlastfoot
% capacity mwp & 100 & 0.69 & 0.83*** & 0.78*** & 0.80 &  \\
GVA by veterinary activities & 93 & 0.41 & 0.64*** & 0.63*** & 0.64 &  \\
2.1. Arable land & 100 & 0.43 & 0.66*** & 0.56*** & 0.61 & 
\cite{van_de_ven_potential_2021} \\
2.1.1. Non-irrigated arable land & 100 & 0.43 & 0.66*** & 0.56*** & 0.61 & \cite{van_de_ven_potential_2021} \\
1.1. Urban fabric & 100 & 0.33 & 0.58*** & 0.62*** & 0.60 & \cite{alghanem_global_2024} \\
1. Artificial Surfaces & 100 & 0.37 & 0.61*** & 0.59*** & 0.60 &  \\
1.1.2. Discontinuous urban fabric & 100 & 0.34 & 0.58*** & 0.62*** & 0.60 &  \\
2. Agricultural Areas & 100 & 0.44 & 0.66*** & 0.53*** & 0.59 & \cite{kruitwagen_global_2021,alghanem_global_2024} \\
1.4.2. Sport and leisure facilities & 100 & 0.31 & 0.55*** & 0.55*** & 0.55 &  \\
1.4. Artificial, non-agricultural vegetated areas & 100 & 0.27 & 0.52*** & 0.55*** & 0.54 &  \\
1.3. Mine, dump and construction sites & 100 & 0.28 & 0.53*** & 0.48*** & 0.51 &  \\
GVA by Production sector & 93 & 0.19 & 0.44*** & 0.55*** & 0.49 &  \\
1.2.4. Airports & 100 & 0.33 & 0.58*** & 0.38*** & 0.48 &  \\
3.1.1. Broad-leaved forest & 100 & 0.21 & 0.45*** & 0.52*** & 0.48 &  \\
GVA by manufacturing & 93 & 0.15 & 0.38*** & 0.55*** & 0.47 & \cite{alghanem_global_2024} \\
GVA by manufacture of electronic, optical and electrical products & 93 & 0.17 & 0.41*** & 0.53*** & 0.47 &  \\
1.3.1. Mineral extraction sites & 100 & 0.22 & 0.47*** & 0.43*** & 0.45 &  \\
Mean sea level pressure & 90 & 0.14 & 0.37*** & 0.52*** & 0.45 &  \\
GHI  & 90 & 0.13 & 0.37*** & 0.51*** & 0.44 & \cite{westacott_novel_2016,yu_deepsolar_2018,aklin_geography_2018,balta-ozkan_regional_2015,celik_review_2020,alghanem_global_2024} \\
GVA by agriculture, forestry and fishing; mining and quarrying & 93 & 0.16 & 0.4*** & 0.46*** & 0.43 & \cite{mayer_deepsolar_2020,alghanem_global_2024} \\
3.1.3. Mixed forest & 100 & 0.20 & 0.45*** & 0.42*** & 0.43 &  \\
GVA by specialised construction activities & 93 & 0.14 & 0.38*** & 0.48*** & 0.43 &  \\
GVA by manufacture of wood and paper products and printing & 93 & 0.14 & 0.38*** & 0.49*** & 0.43 &  \\
1.2. Industrial, commercial and transport units & 100 & 0.20 & 0.44*** & 0.4*** & 0.42 &  \\
Air temperature & 90 & 0.14 & 0.37*** & 0.47*** & 0.42 & \cite{yu_deepsolar_2018,alghanem_global_2024} \\
2.3. Pastures & 100 & 0.18 & 0.43*** & 0.41*** & 0.42 &  \\
GVA by activities of households  & 93 & 0.17 & 0.41*** & 0.43*** & 0.42 &  \\
GVA by residential care activities & 93 & 0.15 & 0.38*** & 0.47*** & 0.42 &  \\
2.3.1. Pastures & 100 & 0.18 & 0.43*** & 0.41*** & 0.42 & \\
1.3.3. Construction sites & 100 & 0.10 & 0.32*** & 0.44*** & 0.38 &  \\
GVA by other personal service activities & 93 & 0.11 & 0.33*** & 0.43*** & 0.38 &  \\
GVA by motor trades  & 93 & 0.05 & 0.22*** & 0.53*** & 0.38 &  \\
GVA by other manufacturing, repair and installation & 93 & 0.07 & 0.27*** & 0.47*** & 0.37 &  \\
2.4. Heterogeneous agricultural areas & 100 & 0.16 & 0.4*** & 0.32*** & 0.36 &  \\
Region area & 100 & 0.09 & 0.29*** & 0.43*** & 0.36 & \cite{collier_distributed_2023,fuentes_solar_2024} \\
GVA by manufacture of machinery and transport equipment & 93 & 0.04 & 0.19*** & 0.51*** & 0.35 &  \\
GVA by owner-occupiers' imputed rental & 93 & 0.10 & 0.32*** & 0.38*** & 0.35 &  \\
GVA by construction & 93 & 0.09 & 0.3*** & 0.4*** & 0.35 &  \\
GVA by manufacture of petroleum, chemicals and other minerals & 93 & 0.04 & 0.21*** & 0.5*** & 0.35 &  \\
GVA by manufacture of basic and fabricated metal products & 93 & 0.04 & 0.19*** & 0.49*** & 0.34 &  \\
2.4.3. Land principally occupied by agriculture, with significant areas of natural vegetation & 100 & 0.14 & 0.38*** & 0.3*** & 0.34 &  \\
1.2.1. Industrial or commercial units & 100 & 0.12 & 0.34*** & 0.34*** & 0.34 &  \\
2.4.2. Complex cultivation patterns & 100 & 0.09 & 0.3*** & 0.33*** & 0.32 &  \\
GVA by wholesale and retail trade; repair of motor vehicles & 93 & 0.04 & 0.21*** & 0.4*** & 0.30 &  \\
GVA by manufacture of food, beverages and tobacco  & 93 & 0.10 & 0.31*** & 0.29*** & 0.30 &  \\
GVA by accommodation & 93 & 0.05 & 0.23*** & 0.34*** & 0.29 &  \\
GVA by social work activities & 93 & 0.05 & 0.23*** & 0.34*** & 0.29 &  \\
GVA by retail trade  & 93 & 0.04 & 0.2*** & 0.38*** & 0.29 &  \\
Wind speed & 90 & 0.09 & 0.3*** & 0.29*** & 0.29 & \cite{yu_deepsolar_2018} \\
GVA by accommodation and food service activities & 93 & 0.04 & 0.19*** & 0.36*** & 0.28 &  \\
4.1.1. Inland marshes & 100 & 0.07 & 0.27*** & 0.28*** & 0.28 &  \\
GVA by civil engineering & 93 & 0.05 & 0.22*** & 0.35*** & 0.28 &  \\
GVA by real estate activities & 93 & 0.05 & 0.22*** & 0.35*** & 0.28 &  \\
GVA by electricity, gas, water; sewerage and waste management & 93 & 0.06 & 0.24*** & 0.31*** & 0.28 &  \\
GVA by other service activities & 93 & 0.02 & 0.13*** & 0.4*** & 0.27 &  \\
GVA by education & 93 & 0.06 & 0.24*** & 0.3*** & 0.27 &  \\
GVA by wholesale trade  & 93 & 0.02 & 0.15*** & 0.38*** & 0.27 &  \\
GVA by postal and courier activities  & 93 & 0.03 & 0.18*** & 0.36*** & 0.27 &  \\
GVA by public administration and defence  & 93 & 0.06 & 0.23*** & 0.29*** & 0.26 &  \\
GVA by human health and social work activities & 93 & 0.05 & 0.22*** & 0.28*** & 0.25 &  \\
GVA by construction of buildings & 93 & 0.03 & 0.17*** & 0.33*** & 0.25 &  \\
2.2.2. Fruit trees and berry plantations & 100 & 0.01 & 0.08*** & 0.4*** & 0.24 &  \\
GVA by food and beverage service activities & 93 & 0.02 & 0.15*** & 0.33*** & 0.24 &  \\
2.2. Permanent crops & 100 & 0.01 & 0.08*** & 0.4*** & 0.24 & \\
1.3.2. Dump sites & 100 & 0.06 & 0.25*** & 0.21*** & 0.23 &  \\
GVA by all industries & 93 & 0.01 & 0.11*** & 0.36*** & 0.23 & \cite{alghanem_global_2024,liu_forecasting_2022} \\
GVA by architectural and engineering activities  & 93 & 0.02 & 0.14*** & 0.31*** & 0.23 &  \\
4.2.1. Salt marshes & 100 & 0.05 & 0.23*** & 0.23*** & 0.23 &  \\
3.1. Forests & 100 & 0.00 & 0.06** & 0.39*** & 0.23 &  \\
2.4.4. Agro-forestry areas & 100 & 0.08 & 0.28*** & 0.13*** & 0.21 &  \\
GVA by other professional, scientific and technical activities & 93 & 0.01 & 0.1*** & 0.3*** & 0.20 &  \\
GVA by services sector& 93 & 0.00 & 0.06** & 0.32*** & 0.19 &  \\
GVA by rental and leasing activities & 93 & 0.01 & 0.08*** & 0.28*** & 0.18 &  \\
GVA by human health activities & 93 & 0.02 & 0.14*** & 0.2*** & 0.17 &  \\
4.2. Marine wetlands & 100 & 0.06 & 0.24*** & 0.09*** & 0.16 &  \\
GVA by real estate activities, excluding imputed rental & 93 & 0.00 & 0.04* & 0.28*** & 0.16 &  \\
GVA by repair of computers, personal and household goods & 93 & 0.00 & 0.06** & 0.24*** & 0.15 &  \\
1.2.2. Road and rail networks and associated land & 100 & 0.01 & 0.1*** & 0.2*** & 0.15 &  \\
GVA by transportation and storage & 93 & 0.01 & 0.07*** & 0.24*** & 0.15 &  \\
GVA by activities of membership organisations & 93 & 0.00 & -0.03 & 0.32*** & 0.15 &  \\
4.1.2. Peat bogs & 100 & 0.01 & -0.1*** & -0.21*** & -0.15 &  \\
GVA by head offices and management consultancy & 93 & 0.00 & -0.01 & 0.3*** & 0.14 &  \\
GVA by services to buildings and landscape activities & 93 & 0.01 & 0.08*** & 0.21*** & 0.14 &  \\
GVA by professional, scientific and technical activities  & 93 & 0.00 & -0.0 & 0.28*** & 0.14 &  \\
3.3.3. Sparsely vegetated areas & 100 & 0.01 & -0.11*** & -0.16*** & -0.14 &  \\
Total precipitation & 100 & 0.00 & -0.04* & -0.23*** & -0.14 & \cite{alghanem_global_2024} \\
GVA by administrative and support service activities & 93 & 0.00 & 0.02 & 0.24*** & 0.13 &  \\
GVA by warehousing and transport support activities & 93 & 0.00 & 0.04 & 0.22*** & 0.13 &  \\
GVA by gambling and betting; sports and recreation activities& 93 & 0.00 & 0.04 & 0.2*** & 0.12 &  \\
4.2.3. Intertidal flats & 100 & 0.03 & 0.18*** & 0.07** & 0.12 &  \\
GVA by land, water and air transport & 93 & 0.00 & 0.04* & 0.2*** & 0.12 &  \\
GVA by office administration and business support activities  & 93 & 0.00 & -0.0 & 0.23*** & 0.12 &  \\
3.3.4. Burnt areas & 100 & 0.01 & -0.07*** & -0.16*** & -0.12 &  \\
3. Forest And Seminatural Areas & 100 & 0.00 & -0.05* & 0.26*** & 0.11 &  \\
GVA by creative, arts, entertainment and cultural activities & 93 & 0.00 & -0.03 & 0.25*** & 0.11 &  \\
5.2.1. Coastal lagoons & 100 & 0.02 & 0.15*** & 0.07*** & 0.11 &  \\
5. Water Bodies & 100 & 0.00 & -0.06** & 0.29*** & 0.11 &  \\
GVA by research and development; advertising and market research & 93 & 0.00 & 0.02 & 0.18*** & 0.10 &  \\
3.3.1. Beaches, dunes, sands & 100 & 0.00 & 0.06** & 0.13*** & 0.10 &  \\
GVA by legal and accounting activities & 93 & 0.00 & -0.06** & 0.26*** & 0.10 &  \\
GVA by telecommunications; information technology & 93 & 0.00 & 0.01 & 0.19*** & 0.10 &  \\
GVA by arts, entertainment and recreation & 93 & 0.00 & -0.0 & 0.2*** & 0.10 &  \\
GVA by manufacture of textiles, wearing apparel and leather & 93 & 0.00 & 0.06** & 0.15*** & 0.10 &  \\
3.1.2. Coniferous forest & 100 & 0.01 & -0.08*** & 0.28*** & 0.10 &  \\
3.3.2. Bare rocks & 100 & 0.01 & -0.09*** & -0.1*** & -0.10 &  \\
5.1.2. Water bodies & 100 & 0.01 & -0.07*** & 0.25*** & 0.09 &  \\
GVA by employment activities; tourism and security services & 93 & 0.00 & -0.03 & 0.2*** & 0.09 &  \\
5.1. Inland waters & 100 & 0.00 & -0.07*** & 0.25*** & 0.09 &  \\
5.2.2. Estuaries & 100 & 0.03 & 0.17*** & -0.01 & 0.08 &  \\
GVA by information and communication & 93 & 0.00 & -0.02 & 0.18*** & 0.08 &  \\
GVA by financial service activities & 93 & 0.01 & -0.07*** & 0.24*** & 0.08 &  \\
GVA by financial and insurance activities & 93 & 0.00 & -0.07** & 0.24*** & 0.08 &  \\
GVA by publishing; film and TV production and broadcasting & 93 & 0.00 & -0.07** & 0.2*** & 0.07 &  \\
GVA by insurance, pension funding and auxiliary financial activities & 93 & 0.00 & -0.06** & 0.21*** & 0.07 &  \\
3.3. Open spaces with little or no vegetation & 100 & 0.01 & -0.1*** & -0.02 & -0.06 &  \\
4.1. Inland wetlands & 100 & 0.01 & -0.1*** & -0.02 & -0.06 &  \\
1.2.3. Port areas & 100 & 0.00 & -0.0 & 0.1*** & 0.05 &  \\
4. Wetlands & 100 & 0.01 & -0.09*** & -0.01 & -0.05 &  \\
1.4.1. Green urban areas & 100 & 0.01 & -0.11*** & 0.06** & -0.03 &  \\
3.2.4. Transitional woodland-scrub & 100 & 0.01 & -0.11*** & 0.15*** & 0.02 &  \\
3.2.2. Moors and heathland & 100 & 0.01 & -0.1*** & 0.06** & -0.02 &  \\
% Year & 100 & 0.00 & 0.0 & -0.04* & -0.02 &  \\
5.1.1. Water courses & 100 & 0.01 & 0.09*** & -0.06** & 0.01 &  \\
3.2. Scrub and/or herbaceous associations & 100 & 0.01 & -0.09*** & 0.11*** & 0.01 &  \\
1.1.1. Continuous urban fabric & 100 & 0.00 & -0.04 & 0.03 & -0.01 &  \\
3.2.1. Natural grassland & 100 & 0.00 & -0.06** & 0.03 & -0.01 &  \\
5.2.3. Sea and ocean & 100 & 0.00 & -0.03 & 0.0 & -0.01 &  \\
% national capacity mwp & 100 & 0.00 & 0.01 & -0.04* & -0.01 &  \\
5.2. Marine waters & 100 & 0.00 & -0.0 & 0.0 & 0.00 &  \\

\end{longtable}

\vspace{5pt} % Adds a bit of space between the table and the note
\noindent\textsuperscript{***}$p<0.001$, 
\textsuperscript{**}$p<0.01$, 
\textsuperscript{*}$p<0.05$

% \multicolumn{3}{l}{\textsuperscript{***}$p<0.001$, 
%   \textsuperscript{**}$p<0.01$, 
%   \textsuperscript{*}$p<0.05$}

\end{landscape}

\section*{Acknowledgements}
Hussah Alghanem would like to thank Imam Abdulrahman Bin Faisal University and the Government of Saudi Arabia for her PhD scholarship. 

\section*{Data Availability}
Datasets and code related to this article are provided in a supplementary file

% \appendix
% \section{Example Appendix Section}


%%       \cite{<label>} ==> [1]

%%
% Example citation, See \cite{lamport94}.

% \bibliographystyle{elsarticle-num} 
% \bibliographystyle{vancouver-authoryear} 
\bibliographystyle{elsarticle-num-names}
\bibliography{bibliography}




% %% For numbered reference style

%   \textit{\LaTeX: a document preparation system},
%   Addison Wesley, Massachusetts,

% \end{thebibliography}
\end{document}